\documentclass[a4paper,11pt]{article}
\usepackage{pos}
\usepackage[T1]{fontenc}
\usepackage{amsmath}
\usepackage{graphicx}
\usepackage{tabularx}
\usepackage{ragged2e}
\usepackage[separate-uncertainty = true,multi-part-units=single]{siunitx}
\usepackage{caption}
\usepackage{subcaption}
\usepackage{float}
\usepackage{comment}
\usepackage{physics}
\usepackage{hyperref}
\usepackage{mathtools}

\RenewCommandCopy\qty\SI
\DeclareSIUnit\barn{b}

\def\ttt#1{\texttt{\small #1}}

\newcommand{\epem}{\mathrm{e}^+\mathrm{e}^-}

\newcommand{\ellell}{\ell^+\ell^-}
\newcommand{\gammaUPC}{\ttt{gamma-UPC}}

\newcommand{\sqrtsnn}{\sqrt{s_{_\text{NN}}}}
\newcommand{\madgraph}{\textsc{MadGraph5\_aMC@NLO}}
\newcommand{\mgshort}{\textsc{MG5\_aMC}}
\newcommand{\helaconia}{\textsc{HELAC-Onia}}
\newcommand{\lhe}{\textsc{lhe}}
\newcommand{\superchic}{\textsc{Superchic}}

\newcommand{\gaga}{\gamma\gamma}

\newcommand{\kt}{k_{\perp}}
\newcommand{\qt}{\vb{q_\perp}}

\title{Improved modeling of photon-photon processes in ultraperipheral collisions at hadron colliders}
\ShortTitle{Improved modeling of $\gaga$ processes in ultraperipheral collisions at hadron colliders}

\author*[a]{Nicolas Crépet}
\author[b]{David d'Enterria}
\author[c]{Hua-Sheng Shao}
\affiliation[a]{Ecole Normale Supérieur Paris-Saclay, Gif-sur-Yvette,France}

\affiliation[b]{EP, CERN, Meyrin, Switzerland}

\affiliation[c]{Laboratoire de Physique Th\'eorique et Hautes Energies (LPTHE), UMR 7589, Sorbonne Universit\'e et CNRS, 4 place Jussieu, 75252 Paris Cedex 05, France}

\emailAdd{nicolas.crepet@ens-paris-saclay.fr}

\abstract{
The CERN LHC is not only the current energy-frontier collider for parton-parton collisions, but has proven a powerful photon collider providing photon-photon ($\gaga$) collisions at center-of-mass energies and luminosities never reached before. The latest theoretical developments implemented in the \gammaUPC\ Monte Carlo (MC) event generator~\cite{Shao:2022cly}, which can calculate arbitrary exclusive final state produced via $\gaga$ fusion in ultraperipheral collisions (UPCs) of protons and/or nuclei at the LHC, are presented. These include azimuthal modulations of dilepton pairs produced in the $\gaga\to\ellell$ process, and neutron emission probabilities for photoexcited lead ions in PbPb UPCs. A few comparisons of the results of the updated \gammaUPC~v.1.6 code to relevant RHIC and LHC data are presented.}

\FullConference{31st International Workshop on Deep Inelastic Scattering (DIS2024)\\
 8–12 April 2024\\
Grenoble, France\\}

\begin{document}
\maketitle
\section{Introduction}
Over the past decade, the CERN Large Hadron Collider (LHC) has been accelerating hadrons to achieve collisions at unprecedented nucleon-nucleon c.m.\ energies (up to $\sqrtsnn = \SI{13.6}{\tera\electronvolt}$) and integrated luminosities (several hundred \unit{\per\femto\barn} per year). Beyond the unique studies of hadronic collisions conducted since 2010, 
the LHC has also studied photon-photon ($\gaga$) collisions at an hitherto unexplored kinematic regime by exploiting the large fluxes of quasireal photons emitted by the accelerated hadrons~\cite{Brodsky:1971ud,Budnev:1975poe}. Such $\gaga$ processes can be studied in particularly clean conditions in the so-called ultraperipheral collisions (UPCs) where the colliding hadrons interact with transverse separations larger than their matter radii, i.e.\ without hadronic overlap, and survive their purely electromagnetic interaction~\cite{Baltz:2009jk}. The phenomenological study of $\gaga$ collisions in UPCs has been significantly facilitated with the recent development of the \gammaUPC\ code~\cite{Shao:2022cly}, which allows the automated computation of arbitrary $\gaga\to\rm X$ processes (including Standard Model, and beyond, final states), in combination with the \madgraph~(\mgshort\ hereafter)~\cite{Alwall:2014hca} or \helaconia~\cite{Shao:2012iz,Shao:2015vga} event generators. 
Using \gammaUPC, photon-fusion processes have been calculated for the first time up to next-to-leading-order (NLO) accuracy in quantum electrodynamics (QED) and/or quantum chromodynamics (QCD)~\cite{Shao:2022cly,AH:2023ewe,AH:2023kor,Shao:2024dmk}.
In these proceedings, we present additional extensions of the \gammaUPC\ code including the effect of the polarization state of the colliding quasireal photons on the azimuthal modulation of dileptons produced in the $\gaga\to\ellell$ process, as well as the probabilities for photoexcitation and subsequent neutron emission of the Pb ions in UPCs at the LHC. Both developments, plus others (\gammaUPC~v.1.6\footnote{Code downloadable from: \href{http://cern.ch/hshao/gammaupc.html}{http://cern.ch/hshao/gammaupc.html}}), will be presented in detail elsewhere~\cite{notrearticle}.

\section{Azimuthal modulation in UPC lepton pair production}

\begin{figure}[htpb!]
    \centering
    \includegraphics[width = 0.8\textwidth]{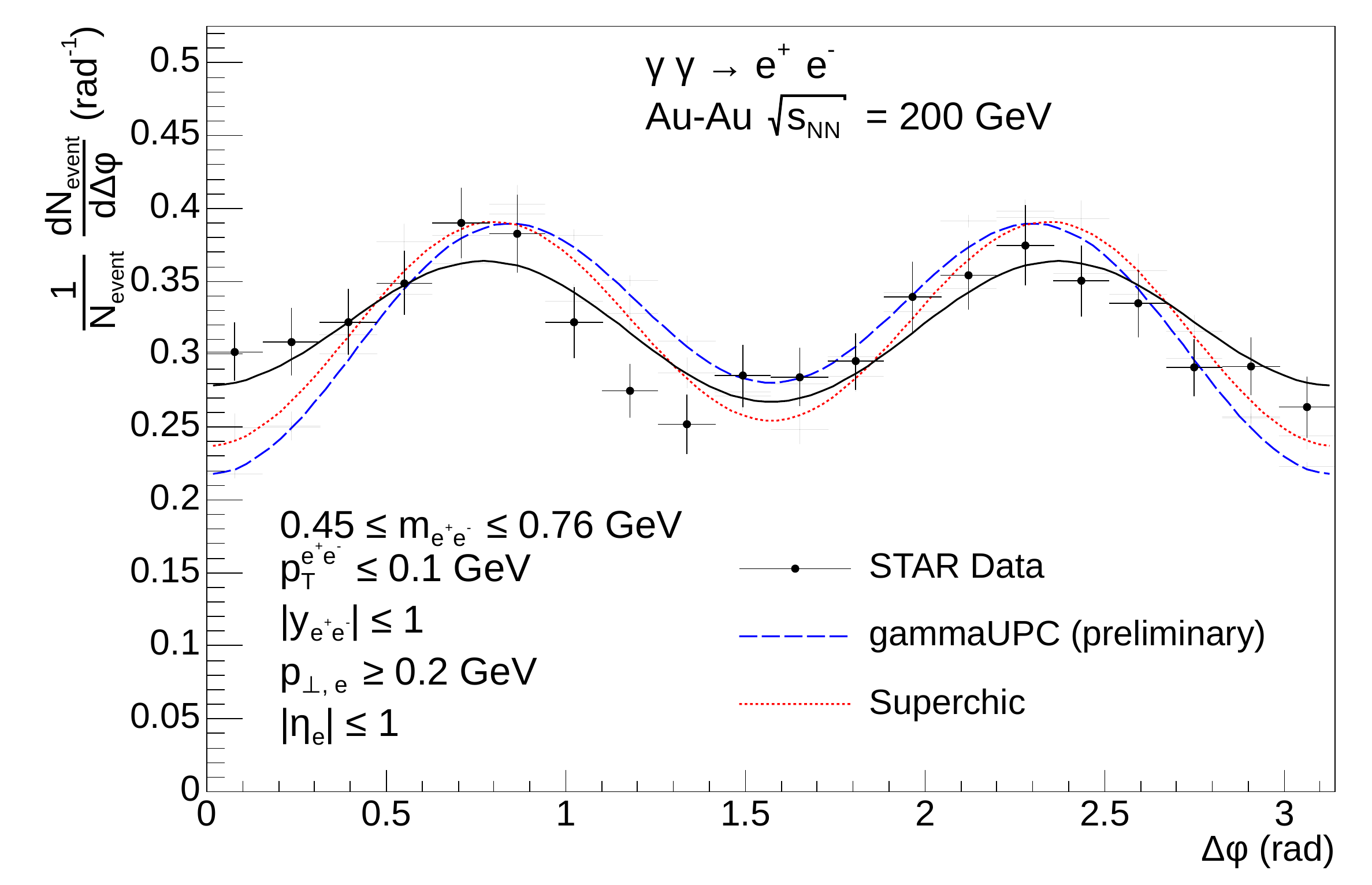}
    \caption{Normalized dielectron $\rm dN/d\Delta\varphi$ distributions for $\gaga \to \epem$ events in Au-Au UPCs at $\sqrtsnn = \qty{200}{\giga\electronvolt}$, with the kinematics cuts listed. The STAR data (points)~\cite{STAR:2019wlg} are compared to our predictions (blue dashed curve) and that of \superchic~(red dotted curve). 
    \label{fig:distrib_deltaphi_XX}}
\end{figure}

In UPCs, the electric field associated with a charge accelerated at high energies vibrates in a single, straight-line direction. This implies that its associated quasireal photon flux, in the equivalent photon approximation (EPA)~\cite{vonWeizsacker:1934nji,Williams:1934ad}, is linearly polarized. One particularly clean way to probe the polarization of photons in UPCs is studying the azimuthal angle distribution of dileptons produced in $\gaga\to\ellell$ processes~\cite{Brandenburg:2022tna}. The differential cross section with respect to the angle $\Delta\varphi$ between $\qt = \vb{k_{1\perp}} + \vb{k_{2\perp}}$ and $\vb{P_\perp} = \frac{\vb{k_{1\perp}} - \vb{k_{2\perp}}}{2}$, where $\vb{k_{1\perp}}$ and $\vb{k_{2\perp}}$ are the transverse momentum of produced leptons, can be decomposed into three terms:
\begin{equation}
\frac{\text{d}\sigma}{\text{d}\Delta\varphi} \propto A + B\cos(2\Delta\varphi) + C\cos(4\Delta\varphi),
\label{eq:distribution}
\end{equation}
where 
$A$, $B$, and $C$ are coefficients (GeV$^{-2}$ units) that can be derived from convolutions of the photon transverse-momentum distribution (TMD)~\cite{Li:2019yzy} (cf. Eqs.~(4--6) therein). 
So far, the combination of \gammaUPC\ with \mgshort\ or \helaconia\ operates within the collinear factorization approach, and the azimuthal modulation represented by Eq.~(\ref{eq:distribution}), properly accounted for by TMD factorization, is absent.
In order to restore the full transverse-momentum dependencies of the photon fluxes, our \gammaUPC\ setup incorporates small extra transverse momentum $\kt$ and azimuthal angle $\varphi$ ``smearings" of the initial photons in events generated within the collinear factorization MC setup. The implementation of the $\kt$ smearing alone has been described in Ref.~\cite{Shao:2024dmk}. 
In these proceedings, we report the new implementation of the simultaneous $\kt$ and $\varphi$ smearing in the \gammaUPC\ framework. The photon TMD coefficients given by Eq.~(\ref{eq:distribution}) have been incorporated into our setup through a smearing of the initial and final states performed by running a {\tt python} script on the MC output Les Houches (\lhe) file that modifies the kinematics of external particles in each event. This implementation has been tested by simulating $\gaga \to \epem$ events in Au-Au UPCs at $\sqrtsnn = \SI{200}{\giga\eV}$ within the fiducial cuts corresponding to the measurement of the STAR collaboration~\cite{STAR:2019wlg}: $m_{\epem} \in [0.45,2.6]$ GeV, $p_{\text{T}}^{\epem} \leq \qty{0.1}{\giga\electronvolt}$, $|y_{\epem}| \leq 1$,  $p_{\perp,\rm e} \geq \qty{0.2}{\giga\electronvolt}$, and $|\eta_{\rm e}| \leq 1$.
Our preliminary result (blue dashed curve) is shown in \autoref{fig:distrib_deltaphi_XX} compared to the azimuthal modulation measured in the experimental data (black symbols), and to the alternative prediction from the \superchic~model (red dotted curve)~\cite{Harland-Lang:2023ohq}. Within uncertainties, both MC predictions can reproduce the modulation observed in data, thereby confirming the linearly polarized nature of the incoming photons.


\section{Coulomb excitation and neutron emission in UPCs with Pb ions}

\begin{figure}
    \centering
    \includegraphics[width=0.53\linewidth]{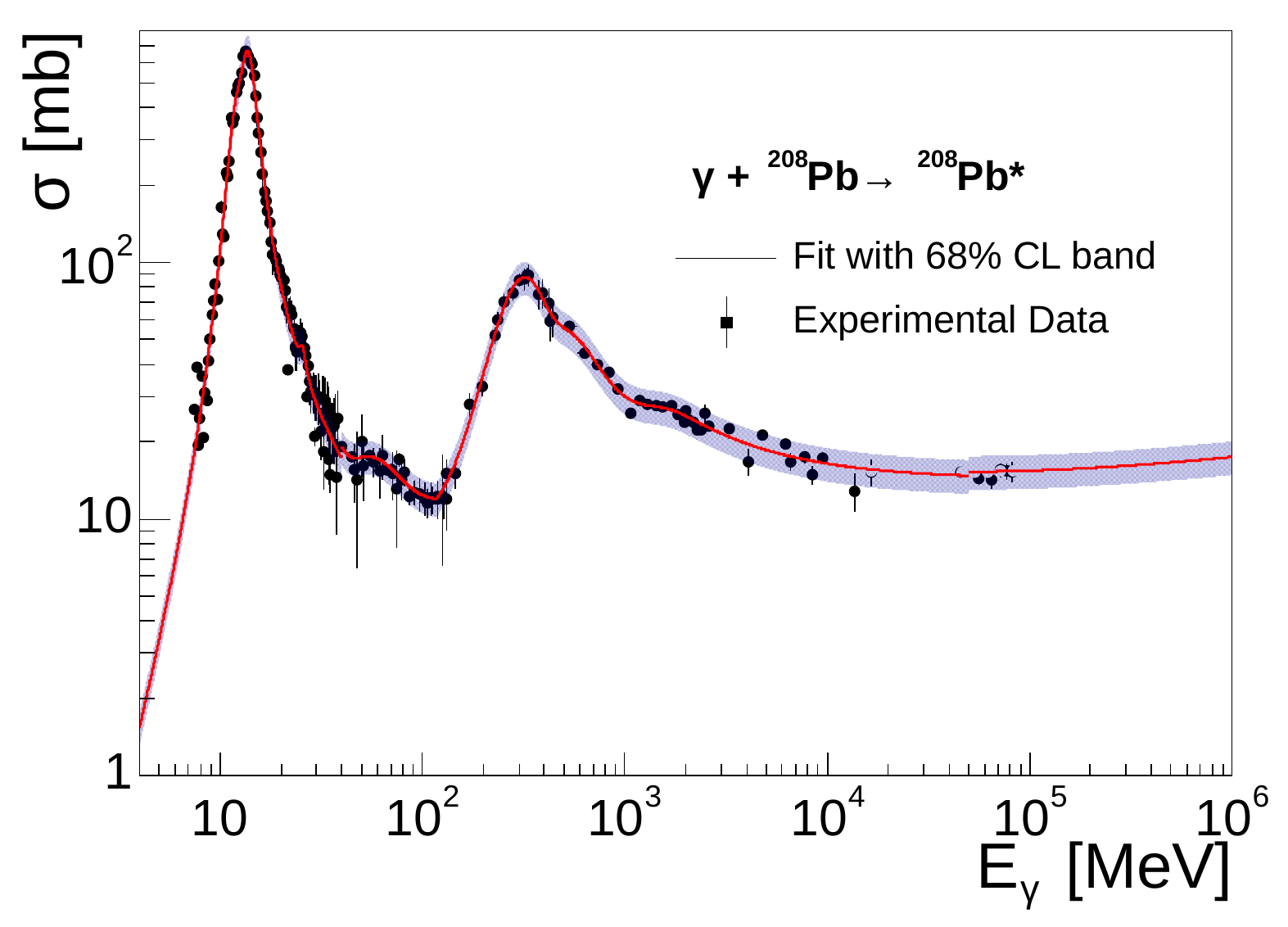}
    \includegraphics[width=0.46\textwidth]{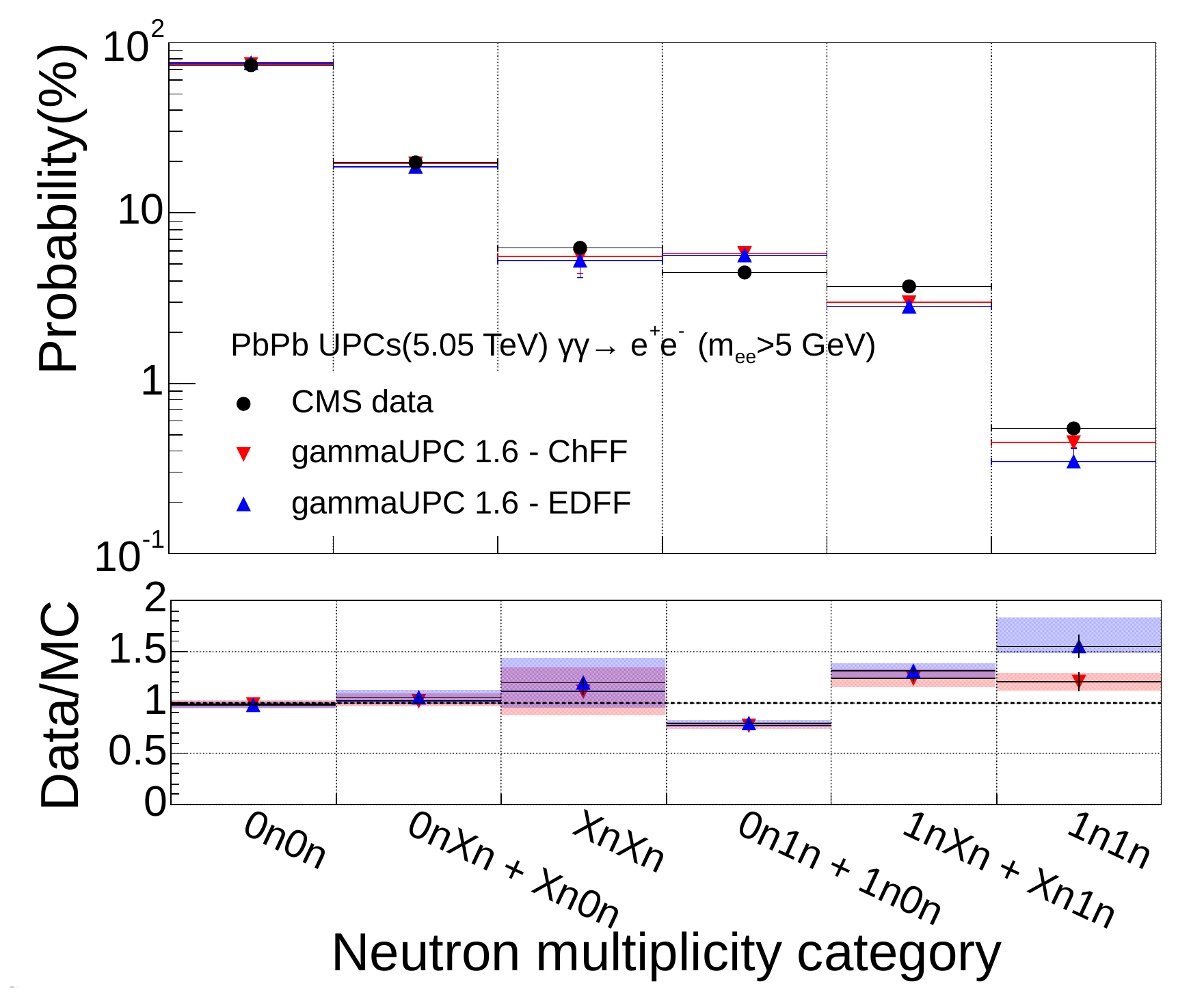}
    \caption{Left: Experimental Pb photoabsorption cross sections as a function of photon energy, $\sigma_{\gamma \rm Pb\to Pb^*}(E_\gamma)$, fitted to our parametrization (red curve with violet band uncertainties). Right: Probability for different neutron emission categories in the $\gaga\to \epem$ process measured by CMS in PbPb UPCs at $\sqrtsnn = 5.02$~TeV~\cite{CMS:2024tfd} compared to our predictions with ChFF and EDFF $\gamma$ fluxes (the bottom panel shows the data/\gammaUPC\ ratio).}
    \label{fig:gammaAbs}
\end{figure}

The second improvement implemented into the \gammaUPC\ code is the calculation of the photoexcitation probability of the nuclei in UPCs, due to soft Coulomb photon exchanges between them taking place simultaneously with the hard $\gaga$ reaction, and their subsequent deexcitation via neutron emission~\cite{Baltz:2009jk}. The neutron(s) emitted from the excited nuclei in UPCs can be detected in very forward (zero-degree) calorimeters, and thereby are commonly used by the experiments to trigger on photon-photon interactions.
The most straightforward way to implement such processes is by adding a probability to emit $X$ and $Y$ neutrons by the photoexcited nuclei A and B (separated by an impact parameter separation $|\pmb{b}_1-\pmb{b}_2|$), $P_{X\mathrm{n}Y\mathrm{n}}(|\pmb{b}_1-\pmb{b}_2|)$, 
that multiplies the no-inelastic hadronic interaction probability, $P_\text{no\,inel}(|\pmb{b}_1-\pmb{b}_2|)$, inside the convolution integral of the two photon fluxes:
\begin{multline}
\!\!\frac{\mathrm{d}^2N^{(\mathrm{AB},X\mathrm{n}Y\mathrm{n})}_{\gamma_1/Z_1,\gamma_2/Z_2}}{\mathrm{d}E_{\gamma_1}\mathrm{d}E_{\gamma_2}} =\!\!\int\!\!\mathrm{d}^2\pmb{b}_1\mathrm{d}^2\pmb{b}_2\,P_{X\mathrm{n}Y\mathrm{n}}(|\pmb{b}_1-\pmb{b}_2|) P_\text{no\,inel}(|\pmb{b}_1-\pmb{b}_2|)
N_{\gamma_1/Z_1}(E_{\gamma_1},\pmb{b}_1)N_{\gamma_2/Z_2}(E_{\gamma_2},\pmb{b}_2).\nonumber 
\end{multline}
The probability term is determined from the experimentally measured values of photoabsorption cross sections followed by neutron emission, $\sigma(\gamma \rm Pb\to Pb^\star \to Pb+Xn)$ with $\mathrm{X}\geq 1$. We have fit the individual cross sections measured in photon-lead interactions for various neutron multiplicities (1n, 2n, 3n, 4n, \dots, and their sum), as a function of the incoming photon energy $E_\gamma$ from threshold (a few MeV) up to 16.4~GeV~\cite{Lepretre:1981tf,Carlos:1984lvc,Muccifora:1998ct,Caldwell:1973bu,Michalowski:1977eg,Gheorghe:2024plx}. Beyond this energy, since few data points are available~\cite{Caldwell:1978ik}, we follow the approach used by other MC generators, such as \textsc{nOOn}~\cite{Broz:2019kpl} and \superchic~\cite{Harland-Lang:2023ohq}, and use a Regge-based parameterization of the total photoabsorption cross sections of the proton at high energy~\cite{ZEUS:2001wan}, scaled by the nuclear mass number $A = 208$ times a shadowing factor of $0.65$ so that the resulting fit matches the high energy Pb photodissociation data~\cite{Caldwell:1978ik}. \autoref{fig:gammaAbs} (left) shows the collected experimental photoabsorption cross sections (black points) with our fit results and assigned uncertainties (red curve with violet band).

With the corresponding neutron deexcitation probabilities implemented in \gammaUPC\ as explained above, we can now compare our predictions to the experimental data measured in PbPb UPCs at $\sqrtsnn = 5.02$~TeV. The results are shown in \autoref{fig:gammaAbs} (right) for various neutron emission probabilities (0n0n, 1n1n, XnXn and combinations, where $\rm X\geq 1$ here) in the $\gaga\to \epem$ ($m_{\epem}>5$~GeV) process, obtained using two different $\gamma$ fluxes (based on the charged (ChFF) and electric dipole (EDFF) form factors) compared to the corresponding CMS data~\cite{CMS:2024tfd}. 
Our ChFF-based results (within the assigned theoretical uncertainties) reproduce well the experimental data as indicated by a data/\gammaUPC\ ratio around unity (red symbols in the bottom panel).


\paragraph*{Acknowledgments.---} Supports from the European Union's Horizon 2020 research and innovation program (grant agreement No.\ 824093, STRONG-2020), the ERC (grant 101041109 `BOSON'), the French ANR (grant ANR-20-CE31-0015, ``PrecisOnium''), and the French LIA FCPPN, are acknowledged. Views and opinions expressed are however those of the authors only and do not necessarily reflect those of the European Union or the European Research Council Executive Agency. 

\bibliographystyle{myutphys}
\bibliography{references.bib}

\end{document}